\documentclass[a4paper]{article}
\usepackage[applemac]{inputenc}
\usepackage{amsmath}
\usepackage{bibentry}
\usepackage{url}
\usepackage[pdftex]{color,graphicx}

\begin{document}
\thispagestyle{empty}
\title{\textbf{SPECTRUM TRADING}\\ \textbf{AN ABSTRACTED BIBLIOGRAPHY}}
\author{Loretta Mastroeni - Maurizio Naldi}
\maketitle
\begin{abstract}
This document contains a bibliographic list of major papers on spectrum trading and their abstracts. The aim of the list is to offer researchers entering this field a fast panorama of the current literature. The list is continually updated on the webpage \url{http://www.disp.uniroma2.it/users/naldi/Ricspt.html}. Omissions and papers suggested for inclusion may be pointed out to the authors through e-mail (\textit{naldi@disp.uniroma2.it}).
\end{abstract}
\newpage
\section*{\center{Abstracted Bibliography}}
\bibliographystyle{plain}
\nobibliography{Lic.bib}
\begin{itemize}
\item \bibentry{Aky06} 
\begin{quotation}
\footnotesize{\textit{Abstract}: Today’s wireless networks are characterized by a ﬁxed spectrum assignment policy. However, a large portion of the assigned spectrum is used sporadically and geographical variations in the utilization of assigned spectrum ranges from 15\% to 85\% with a high variance in time. The limited available spectrum and the inefficiency in the spectrum usage necessitate a new communication paradigm to exploit the existing wireless spectrum opportunistically. This new networking paradigm is referred to as NeXt Generation (xG) Networks as well as Dynamic Spectrum Access (DSA) and cognitive 
radio networks. The term xG networks is used throughout the paper. The novel functionalities and current research challenges of the xG networks are explained in detail. More speciﬁcally, a brief overview of the cognitive radio technology is provided and the xG network architecture is introduced. Moreover, the xG network functions such as spectrum management, spectrum mobility and spectrum sharing are explained in detail. The inﬂuence of these functions on the performance of the upper layer protocols such as routing and transport are investigated and open research issues in these areas are also outlined. Finally, the cross-layer design challenges in xG networks are discussed.}
\end{quotation}
\item \bibentry{Aky09}
\begin{quotation}
\footnotesize{\textit{Abstract}: The problem of spectrum scarcity and inefficiency in spectrum usage will be addressed by the newly emerging cognitive radio paradigm that allows radios to opportunistically transmit in the vacant portions of the spectrum already assigned to licensed users. For this, the ability for spectrum sensing, spectrum sharing, choosing the best spectrum among the available options, and dynamically adapting transmission parameters based on the activity of the licensed spectrum owners must be integrated within cognitive radio users. Specifically in cognitive radio ad hoc networks, distributed multihop architecture, node mobility, and spatio-temporal variance in spectrum availability are some of the key distinguishing factors. In this article the important features of CRAHNs are presented, along with the design approaches and research challenges that must be addressed. Spectrum management in CRAHNs comprises spectrum sensing, sharing, decision, and mobility. In this article each of these functions are described in detail from the viewpoint of multihop infrastructureless networks requiring cooperation among users.}
\end{quotation}
\item \bibentry{Bae}
\begin{quotation}
\footnotesize{\textit{Abstract}: It has been widely recognized that the current under-utilization of spectrum across many bands could be alleviated through the application of spectrum markets. So far, discussions of market mechanisms for spectrum allocations have focused primarily on secondary markets, which are managed by licensees. Here we explore the consequences of lifting current restrictions on allocations and ownership, and allowing more extensive markets for allocating spectrum across locations, times, and diverse sets of applications (e.g., broadcast, cellular, broadband data, emergency, etc). To motivate our discussion we first estimate the achievable rate per user that could be provided by sharing a large portion of the spectrum suitable for cellular and broadcast types of services. Our results suggest that in general the demand for spectrum may exceed supply implying that market mechanisms are needed to avoid a tragedy of the commons (i.e., associated with an alternative commons model). We then discuss a two- tier spectrum market structure for wireless services in which licenses for spectrum assets at particular locations are traded as commodities. Spectrum owners can choose to rent or lease their spectrum assets via spot markets at particular locations. Such an approach may lower barriers to entry into the wireless services market thereby facilitating competition and the introduction of new services.}
\end{quotation}

\item \bibentry{Ballon}
\begin{quotation}
\footnotesize{\textit{Abstract}: The concept of flexible spectrum is often considered as a medium-to-long-term solution to overcome some of the current inefficiencies and high entry barriers plaguing the mobile industry. Increasingly, a cognitive pilot channel (CPC) is regarded as a central enabler for flexible spectrum. This paper outlines the CPC concept from a business point of view and clarifies its current status in the standardization and regulation fields. The idea of a worldwide CPC will be under consideration by the World Radio Conference in 2011. Based on several potential CPC implementations, the paper identifies a number of flexible spectrum business configurations and revenue sharing models. It also performs an initial forward-looking evaluation of these models using a business model scorecard approach, and finds that while the scope appears to be limited for a fully competitive, cross-operator spectrum market, several platform models (e.g. association or consortium models) stand out as feasible options.}
\end{quotation}

\item \bibentry{Bayrak}
\begin{quotation}
\footnotesize{\textit{Abstract}: We build a model where consumers have preferences, over a range of varieties, for communications devices and device qualities. Devices use radio spectrum and are produced and sold in a competitive device market. In such a model, the choice on the spectrum management regime greatly affects market outcomes and consumer welfare. The choice on the spectrum management regime effectively determines the final number of firms in the market and the electromagnetic noise levels firms become subject to when deciding on the information transmission capabilities of their devices. We simulate such a model over a wide range of model parameters. We find that the commons regime where the access to and the use of spectrum is not restricted to license holders creates consumer welfare for a wider and more reasonable model parameters than the licensing regime where the access to and the use of spectrum is granted to predetermined number of licensees.}
\end{quotation}

\item \bibentry{Bazelon}
\begin{quotation}
\footnotesize{\textit{Abstract}: At present, no existing market mechanism allows for the trading of radio spectrum between licensed and unlicensed uses. Whenever spectrum is made available for reallocation, the FCC faces a dilemma in determining which access regime to use. For example, the television white spaces are largely unused and available for reallocation. Since both licensed and unlicensed allocations are valuable, allocation decisions (for the TV white spaces or any new band of radio spectrum) must be based on a clear understanding of the trade-offs between the two choices. This article defines economic criteria that can be used in making these important decisions. Economic criteria can go beyond the simple measures of profit and consumer surplus from market transactions. Although some measures of benefit, such as the value of innovation, may be difficult to quantify, the analytic economic framework presented here can easily incorporate them. This analysis does not address any noneconomic considerations in choosing between licensed and unlicensed uses. As one example, the issue of potential societal benefits from promoting minority ownership of spectrum through restricted licenses - something only possible in a licensed regime - is not addressed in this economic analysis. The analysis herein provides the economic information needed for policy analysis; it need not be the sum total of that policy analysis. Standard economic theory tells us that the value of an additional unit of spectrum is equal to the increase in socially beneficial services it produces. For licensed spectrum allowed to trade in markets, this value is relatively easy to calculate: It is the price firms pay for the licensed spectrum. The equation is more complex, however, when unlicensed spectrum is involved. The current value of unlicensed spectrum bands is equal to the sum of the value of the spectrum in all uses in those bands. The incremental value of additional spectrum allocated to unlicensed uses, however, is based - on the relief to congestion the additional spectrum will provide. Unlicensed spectrum also contains a value associated with the possibility of future innovation made available by the lower transaction .costs of gaining access to unlicensed spectrum. This option value increases with additional allocations of unlicensed spectrum, leading to the benefit of incremental option value from additional unlicensed spectrum. The formula for the benefits from additional unlicensed spectrum allocations can be summarized as "congestion alleviation plus incremental option value." I apply the analysis developed in this article to the case of TV white spaces. I use information from the recent auction of the lower 700 MHz band E block to calculate the incremental value of licensing the white spaces. I also calibrate an estimate of the incremental value of the white spaces under an unlicensed allocation. Initial calibration of the economic criteria that determine the trade-off between incremental licensed and unlicensed spectrum allocations indicates that currently licensing incremental allocations is the favored policy. If policy makers choose to allocate incremental spectrum as unlicensed, they should recognize the economic costs of that choice.}
\end{quotation}

\item \bibentry{Borgers}
\begin{quotation}
\footnotesize{\textit{Abstract}: This paper analyses the procedures used by different European countries for awarding spectrum licences to potential operators of third generation (3G) mobile telephone networks. We contrast market-based methods, such as auctions, with bureaucratic methods, such as ‘beauty contests’. They have been used for decisions about two major questions: (1) How many licences should be awarded, and how much spectrum should each licence give access to?; (2) Which companies should receive which licences, and how much should they pay for their licences? Most countries used a bureaucratic process to answer the ﬁrst question. However, Germany, Austria and Greece were different, and constructed auctions in which the number and size of licences were determined by the auction itself. As for the second question, there was 
much variation between countries, and both auctions, and ‘beauty contests’ were 
popular methods. We have four main ﬁndings. First, the bureaucratic procedure 
used by most countries to answer the ﬁrst question led to companies concealing 
relevant information from the authorities. Second, while ﬁrms may have tried to 
manipulate the procedures used in Germany, Austria and Greece to deter entry to 
their markets, they were surprisingly unsuccessful in this. Third, the traditional economic criticisms of beauty contests seem to apply to some, but not to all those that were conducted. Finally, the bidding strategies adopted by the telecom companies were often more complex than those predicted by economic theory}
\end{quotation}

\item \bibentry{Bykowsky03}
\begin{quotation}
\footnotesize{\textit{Abstract}: The development of a successful secondary market for the trading of spectrum is not a foregone conclusion. The multi-dimensional nature of radio spectrum, which requires that a bid to buy and an offer to sell conform across the multiple dimensions, suggests that the market may be very “thin.” In addition, existing commercial users of spectrum have little incentive to sell excess spectrum if such spectrum will be employed by the buyer to provide a service that competes with the service provider by the seller. This paper discusses several steps to enhance market liquidity. One approach involves obtaining participation from federal spectrum users. Another step involves developing a market that both enhances market liquidity and provides participants the opportunity to incorporate a call option in the traded asset.}
\end{quotation}

\item \bibentry{Bykowsky09}
\begin{quotation}
\footnotesize{\textit{Abstract}: This paper is concerned with the merits of employing market forces to address the issues of wireless spectrum congestion and the allocation of spectrum between firms seeking licensed and unlicensed spectrum rights. We show that when unlicensed spectrum is assigned to all competing users during periods of excess demand an inefficient outcome related to the “Tragedy of the Commons” is likely to result. This inefficiency can be substantially reduced when the assignment of users to unlicensed spectrum is based on the bandwidth and latency tolerance needs of the competing users. Further efficiency gains can also occur when users are required to bid to have their “unlicensed spectrum” needs met in the presence of congestion. The paper also examines the merits of creating an auction based market in which firms providing spectrum based services to users bid to have their “spectrum regime” needs satisfied. The objective of this approach is to reduce the incentive that service operators have to misstate their expressed value for a given license regime. The efficiency of this approach is based in large part on the auction mechanism’s ability to solve a “collective action problem” in which firms desiring unlicensed spectrum have an incentive to “free-ride” on the bidding behavior of other unlicensed firms. Together our results open up the possibility that a wide variety of spectrum policy issues may be efficiently solved using a market-based approach. They further suggest that there may be a “hybrid” regime that combines the best features of the license and unlicensed regimes and, thus, lead to a more efficient use of spectrum at any moment in time.}
\end{quotation}

\item \bibentry{Caicedo}
\begin{quotation}
\footnotesize{\textit{Abstract}: The assignment of spectrum licenses through spectrum trading markets can be used as a mechanism to grant access to spectrum to those who value it most and can use it more efficiently. Although various methods for improved spectrum assignment have been proposed along the lines of Dynamic Spectrum Allocation (DSA) and others, several issues related to their implementation are still unsolved. Among them, how to deal with interference in a multi-provider environment and determining the elements and architectures for feasible implementations of spectrum trading markets. This paper analyzes several proposed kinds of trading interactions that may arise in a spectrum trading market and proposes a taxonomy of architectures that could be used to implement them. We also discuss the benefits and limitations of using a single or restricted set of wireless standards in the implementation of a spectrum trading infrastructure as a step towards Wireless Bandwidth Trading which is proposed as a possible path for a realizable instance of markets based on spectrum trading concepts.}
\end{quotation}

\item \bibentry{Cave}
\begin{quotation}
\footnotesize{\textit{Abstract}: The introduction of spectrum trading creates opportunities for operators, singly or jointly, to foreclose entry into downstream markets by accumulating unneeded spectrum holdings. After considering how these issues are treated under administrative methods of spectrum management, the paper examines the degree of substitutability of frequencies with or without regulatory constraints, concluding that the latter are a major source of limitations on substitutability. This may create a case for intervention in the transition to a spectrum market. Alternative forms of intervention are considered, including caps on spectrum holdings or on the acquisition of spectrum at any award}
\end{quotation}

\item \bibentry{Chapin-1}
\begin{quotation}
\footnotesize{\textit{Abstract}: Rapid progress is being made in the technology for dynamic spectrum access (DSA) radio systems. However, the structure and dynamics of the wireless service market must also evolve for DSA to succeed. This article examines the interlinked tecnical and economic issues associated with markets for DSA-based wireless services. We use this analysis to make technical and policy recommendations supporting the commercial success of DSA technology.}
\end{quotation}
\item \bibentry{Chapin-2}
\begin{quotation}
\footnotesize{\textit{Abstract}: A time-limited lease is a set of rights that expires after a specified duration. We analyze ways to use leases to facilitate innovation in radio devices and wireless communication. In our vision, manufacturers include in their devices a simple, secure subsystem that contains a clock and disables specific transmit capabilities if no extension message is received by the end of the lease period. When devices provide this support, regulators may use certification leases rather than permanent grants to accelerate deployment of innovative radios. Spectrum rights holders may use leases to reduce risk in secondary spectrum market transactions. Firms collaborating in innovative wireless service business models can better retain control of their respective rights. We examine both the technical and policy issues associated with leases.}
\end{quotation}

\item \bibentry{Chin-1}
\begin{quotation}
\footnotesize{\textit{Abstract}: Radio resource management (RRM) is one of the most challenging and one of the most important aspects in the provisioning of quality of service (QoS) for wireless communication systems. With the ever increasing size of wireless mobile community and its demand for high-speed multimedia communications, efficient resource management becomes a paramount importance due to limited resources available such as spectrum and power availability. As radio spectrum is a finite resource, new approaches are therefore needed to maximize the existing spectrum to ensure the wireless userspsila and network providerspsila QoS requirements are met. In this paper we propose a novel radio spectrum trading via derivatives contracts as a means to address short-term demands for spectrum whilst ensuring end-to-end QoS is met.}
\end{quotation}

\item \bibentry{Chin-2}
\begin{quotation}
\footnotesize{\textit{Abstract}: The growth in the deployment of wireless systems and technologies shows that in this rapidly evolving and expanding environment, radio resource management (RRM) is fast becoming the most significant aspect in the provisioning of quality of service (QoS) for wireless networks. As radio spectrum is a finite resource, new approaches are therefore needed to maximize the existing spectrum to ensure the wireless users' and network providers' QoS requirements are met. In this paper we propose a novel radio spectrum trading via derivatives contracts as a means to address short-term demands for spectrum whilst ensuring end-to-end QoS is met.}
\end{quotation}

\item \bibentry{Crocioni}
\begin{quotation}
\footnotesize{\textit{Abstract}: The debate on spectrum reforms has mostly focused on the choice between a property rights and a commons regime. This article argues that moving to a property right system requires careful attention to details in order to avoid that ''micro'' rather than ''macro'' factors may prevent efficient trades from taking place. It provides a framework to conduct this assessment, identifies a number of possible concerns and puts forward some solutions.}
\end{quotation}

\item \bibentry{Daoud-1}
\begin{quotation}
\footnotesize{\textit{Abstract}: We study secondary pricing of spectrum in wireless cellular networks employing CDMA at the physical layer. We consider a primary license holder who aims to lease its spectrum within a certain geographic subregion of its own network. Such a transaction has two contrasting economic implications for the seller: On the one hand the seller obtains a revenue due to the exercised price, or rent, of the region. On the other hand, the seller incurs a cost due to (i) reduced spatial coverage of its network and (ii) possible interference from the leased region into the retained portion of its network. We formulate an optimization problem with the objective of profit maximization, and characterize its solutions based on a reduced load approximation that can be shown to be asymptotically exact. The form of optimal prices suggests charging the buyer per admitted call, in proportion with the interference it generates. The charged amount balances the corresponding loss of revenue incurred by the seller due to the influence of an admitted call. We numerically argue that this pricing approach yields better profit compared to some other simplistic techniques.}
\end{quotation}

\item \bibentry{Daoud-2}
\begin{quotation}
\footnotesize{\textit{Abstract}: Optimal price of spectrum in secondary markets is studied. We consider a primary license holder who aims to lease the right to provide service in a given subset of its coverage area. Such a transaction has two contrasting economic implications for the seller: on the one hand the seller obtains a revenue due to the exercised price of the region. On the other hand, the seller incurs a cost due to (i) reduced spatial coverage of its network and (ii) possible interference from the leased region into the retained portion of its network. We formulate an optimization problem with the objective of profit maximization, and characterize its solutions based on a reduced load approximation. The form of optimal price suggests charging each admitted call in proportion to the attendant revenue loss due to the generated interference.}
\end{quotation}

\item \bibentry{Daoud-3}
\begin{quotation}
\footnotesize{\textit{Abstract}: We develop analytical models to characterize pricing of spectrum rights in cellular CDMA networks. Specifically, we consider a primary license holder that aims to lease its spectrum within a certain geographic subregion of its network. Such a transaction has two contrasting economic implications: On the one hand the lessor obtains a revenue due to the exercised price of the region. On the other hand, it incurs a cost due to: (1) reduced spatial coverage of its network; and (2) possible interference from the leased region into the retained portion of its network, leading to increased call blocking. We formulate this tradeoff as an optimization problem, with the objective of profit maximization. We consider a range of pricing philosophies and derive near-optimal solutions that are based on a reduced load approximation (RLA) for estimating blocking probabilities. The form of these prices suggests charging the lessee in proportion to the fraction of admitted calls. We also exploit the special structure of the solutions to devise an efficient iterative procedure for computing prices. We present numerical results that demonstrate superiority of the proposed strategy over several alternative strategies. The results emphasize importance of effective pricing strategies in bringing secondary markets to full realization.}
\end{quotation}

\item \bibentry{Daoud-4}
\begin{quotation}
\footnotesize{\textit{Abstract}: We consider pricing secondary access to wireless spectrum in cellular CDMA networks. We study the case for a primary license holder interested in leasing the right of providing service in a given geographical region of its coverage network. The goal is to price access to the cells in that region under heterogeneous call traffic demand with the objective of profit maximization. While a revenue is gained from the leased region due to the exercised price, the primary license holder incurs a loss due to reduced spatial coverage of the network and also due to interference effect from the leased into the retained region. We exploit the spatial effect of interference due to geographical locations of the cells and set a price per cell rather than pricing the whole region by a scalar quantity. We employ reduced load approximations which have proved useful in classical telephony and characterize optimal prices for different pricing philosophies, e.g., flat pricing and demand-based pricing. The obtained formula of prices suggests charging per admitted call in proportion with the interference that the call generates. The charged amount balances the corresponding loss of revenue due to the influence of an admitted call. We present an iterative price computing technique and provide a numerical study in support of our analytical results.}
\end{quotation}

\item \bibentry{Vany69}
\item \bibentry{Vany98}
\begin{quotation}
\footnotesize{\textit{Abstract}: The spectrum auctions were a step toward the Herzel?Coase vision of a flexible and efficient market for spectrum. This article examines what remains to be done. Spectrum must be unbundled from broadcast and transmission facilities. The “commoditization” of spectrum will facilitate standardization, price discovery, and open access to diverse users. A liquid secondary spectrum market will lower transactions and entry cost, making telecommunications markets contestable. Auctions should be used to elicit a supply of spectrum from licensees as well as to allocate it to new users. In closing the spectrum commons, Congress granted use to a privileged few. Unbundled spectrum property rights, commoditization, and open markets will give the public access to this public resource.}
\end{quotation}

\item \bibentry{Doyle}
\begin{quotation}
\footnotesize{\textit{Abstract}: This paper focuses on highly fluid markets for trading exclusive spectrum usage-rights. The purpose of the paper is to underline the need for flexible usage-rights policies, as a core facilitator of such markets as well as to stress the need for a greater technical input to the debate. The paper builds on current work in the field of spectrum property rights and exclusive usage-rights. The first half of the paper captures the current state-of-play and presents a framework for visualizing the concepts involved. The paper goes on to make a clear distinction between the defining of a set of exclusive usage-rights and the exercising of those rights. This leads to a discussion of policies that are not alone about defining metrics and setting their desired values but are also about behaviors that involve negotiation and interaction. Through-out the paper the evolution of technology and its affect on the progress towards the goal of fluid spectrum markets is emphasized, as is the need for a very multifaceted approach to the challenges involved.}
\end{quotation}

\item \bibentry{Duan}
\begin{quotation}
\footnotesize{\textit{Abstract}: Dynamic spectrum leasing can greatly enhance the 
spectrum efﬁciency and encourage more ﬂexible services in the 
spectrum market. This paper presents a comprehensive study of 
the interactions of two competitive secondary network operators 
who need to make optimal investment and pricing decisions with 
heterogeneous investment costs and users’ wireless characteris- 
tics. The two operators need to determine their optimal spectrum 
leasing amounts from the spectrum owners, and compete to 
sell the spectrum to secondary users. The asymmetric leasing 
costs and the heterogeneity of users’ channel conditions and 
transmission powers make the analysis quite challenging. We 
model the interactions between the operators and the users as 
a three-stage dynamic game. We show that when the leasing 
costs are small for both operators, there are inﬁnitely many 
investment (leasing) equilibria. When the two costs are high 
and comparable, there exists a unique investment equilibrium 
where both operators lease positive amounts. When the costs 
are high and very different, the operator with the smaller cost 
will dominate the market as a monopolist. We also show that a 
meaningful pricing equilibrium exists only when total bandwidth 
investment of the operators is no larger than a threshold. Each 
user receives a fair resource allocation that only depends on 
the leasing costs of the operators and is independent of other 
users’ wireless characteristics. We also compare the competitive 
equilibria with the case where the two operators cooperate to 
maximize the total proﬁt. We show that the Price of Anarchy 
for the total proﬁt is 75
competition is no larger than 25\% in the worst case. We also 
show that the users always beneﬁt from competition in terms of 
their payoffs.}
\end{quotation}

\item \bibentry{Farquhar}
\begin{quotation}
\footnotesize{\textit{Abstract}: The FCC's historic “command-and-control” approach to spectrum management has contributed to spectrum scarcity, especially as it relates to spectrum below 3 GHz. This approach has also made it difficult for spectrum users to adjust their business plans to the needs of the marketplace. A more flexible regulatory regime would increase spectrum efficiency and foster innovation and new services.
In order to promote more flexible use of spectrum, the authors recommend a number of regulatory and statutory changes, including (1) the elimination of use restrictions for new wireless allocations; (2) the replacement of existing use restrictions with power limits sufficient to minimize the potential for harmful interference; (3) the enactment of rules expressly allowing private parties to “contract around” established interference limits; and (4) the identification of “safe harbor” spectrum leasing arrangements that are deemed permissible under the FCC's license transfer of control requirements.
The authors also recommend a number of statutory reforms, including granting the FCC express statutory authority to auction spectrum licenses held by private parties and assess spectrum fees upon users of auction-exempt spectrum.}
\end{quotation}

\item \bibentry{Freyens}
\begin{quotation}
\footnotesize{\textit{Abstract}: Recent reforms to radio spectrum regulation have sparked controversy over the respective merits of two mutually exclusive liberalization regimes: property rights and commons. This debate is restrictive because it is largely incomplete and misunderstood. It is also costly in terms of opposition and delays to reforms. Goals of efficient spectrum allocation are better served by a wider policy toolkit, inclusive of hybrid and intermediary regimes. In this article I sketch the contours of a ‘spectrum of spectrum regimes’, triangulating regulatory, private ownership and unlicensed approaches. I illustrate this triangulated model, which I then apply to confront allocative decisions in digital dividend policy, such as the FCC’s open access clause in the 700 MHz auction and Ofcom’s current review of UHF spectrum release in the 800 MHz band.}
\end{quotation}

\item \bibentry{Gandhi}
\begin{quotation}
\footnotesize{\textit{Abstract}: In this paper, we propose a low-complexity auction framework to distribute spectrum in real-time among a large number of wireless users with dynamic traffic. Our design consists of a compact and highly expressive bidding format, two pricing models to control tradeoffs between revenue and fairness, and fast auction clearing algorithms to achieve conflict-free spectrum allocations that maximize auction revenue. We develop analytical bounds on algorithm performance and complexity to verify the efficiency of the proposed approach. We also use both simulated and real deployment traces to evaluate the auction framework. We conclude that pricing models and bidding behaviors have significant impact on auction outcomes and spectrum utilization. Any efficient spectrum auction system must consider demand and spectrum availability in local regions to maximize system-wide revenue and spectrum utilization.}
\end{quotation}

\item \bibentry{Glover}
\begin{quotation}
\footnotesize{\textit{Abstract}: Recent technological progress has brought new life to the notion of "just-in-time" spectrum management, raising the concern that further development is being held back by archaic regulatory restrictions imposed by spectrum managers. In the search for a solution, it has become clear that several interdisciplinary challenges still undermine the concepts of Dynamic Spectrum Access (DSA) and Dynamic Spectrum Management (DSM). In this paper we take a step back from the technical detail of implementation and focus on identifying the key milestones to the success of DSM. In identifying a roadmap for DSM, we look at the history of spectrum regulation, charting the key events in the transformation from free and open spectrum into the tightly regulated system of "Command \& Control" that exists today. We outline the similarities between DSM and open spectrum and build on these to present the timeline for spectrum management in terms of a spectrum spiral. We expand on this roadmap, breaking down the key drivers and barriers at each milestone. We conclude that without spectrum scarcity, the concept of DSM is meaningless. But, given that prime spectrum is thought to be scarce, the key remaining barrier to the success of DSM is that of commercial viability.}
\end{quotation}

\item \bibentry{Huang}
\begin{quotation}
\footnotesize{\textit{Abstract}: We study auction mechanisms for sharing spectrum among a group of users, subject to a constraint on the interference temperature at a measurement point. The users access the channel using spread spectrum signaling and so interfere with each other. Each user receives a utility that is a function of the received signal-to-interference plus noise ratio. We propose two auction mechanisms for allocating the received power. The first is an auction in which users are charged for received SINR, which, when combined with logarithmic utilities, leads to a weighted max-min fair SINR allocation. The second is an auction in which users are charged for power, which maximizes the total utility when the bandwidth is large enough and the receivers are co-located. Both auction mechanisms are shown to be socially optimal for a limiting “large system” with co-located receivers, where bandwidth, power and the number of users are increased in fixed proportion. We also formulate an iterative and distributed bid updating algorithm, and specify conditions under which this algorithm converges globally to the Nash equilibrium of the auction.}
\end{quotation}

\item \bibentry{Huang-2}
\begin{quotation}
\footnotesize{\textit{Abstract}: This paper presents a comprehensive analytical study of two competitive cognitive operators' spectrum leasing and pricing strategies, taking into account operators' heterogeneity in leasing costs and users' heterogeneity in transmission power and channel conditions. We model the interactions between operators and users as a three-stage dynamic game, where operators make simultaneous spectrum leasing and pricing decisions in Stages I and II, and users make purchase decisions in Stage III. Using backward induction, we are able to completely characterize the game's equilibria. We show that both operators make the equilibrium leasing and pricing decisions based on simple threshold policies. Moreover, two operators always choose the same equilibrium price despite their difference in leasing costs. Each user receives the same signal-to-noise-ratio (SNR) at the equilibrium, and the obtained payoff is linear in its transmission power and channel gain. We also compare the duopoly equilibrium with the coordinated case where two operators cooperate to maximize their total profit. We show that the maximum loss of total profit due to operators' competition is no larger than 25\%. The users, however, always benefit from operators' competition in terms of their payoffs. We show that most of these insights are robust in the general SNR regime.}
\end{quotation}

\item \bibentry{Ileri}
\begin{quotation}
\footnotesize{\textit{Abstract}: In this paper we develop a framework for competition of future operators likely to operate in a mixed commons/property-rights regime under the regulation of a spectrum policy server (SPS). The operators dynamically compete for customers as well as portions of available spectrum. The operators are charged by the SPS for the amount of bandwidth they use in their services. Through demand responsive pricing, the operators try to come up with convincing service offers for the customers, while trying to maximize their profits. We first consider a single-user system as an illustrative example. We formulate the competition between the operators as a non-cooperative game and propose an SPS-based iterative bidding scheme that results in a Nash equilibrium of the game. Numerical results suggest that, competition increases the user's (customer's) acceptance probability of the offered service, while reducing the profits achieved by the operators. It is also observed that as the cost of unit bandwidth increases relative to the cost of unit infrastructure (fixed cost), the operator with superior technology (higher fixed cost) becomes more competitive. We then extend the framework to a multiuser setting where the operators are competing for a number of users at once. We propose an SPS-based bandwidth allocation scheme in which the SPS optimally allocates bandwidth portions for each user-operator session to maximize its overall expected revenue resulting from the operator payments. Comparison of the performance of this scheme to one in which the bandwidth is equally shared between the user-operator pairs reveals that such an SPS-based scheme improves the user acceptance probabilities and the bandwidth utilization in multiuser systems}
\end{quotation}

\item \bibentry{Ileri-2}
\begin{quotation}
\footnotesize{\textit{Abstract}: In this paper we develop a framework for operator competition based on short term ownership of spectrum resources and devise a dynamic spectrum access method called "Dynamic Property Rights Spectrum Access (D-Pass)". In the D-Pass model, spectrum portions are allocated to operators on a short term basis (few sessions) by a SPS (Spectrum Policy Server) that serves as a controller/enforcer as well as a clearinghouse for spectrum allocations. Prior to each short term allocation, the SPS optimally determines a specific partition of spectrum resources among the operators to maximize a system related objective function. The operators are charged by the SPS for the amount of spectrum they are allocated. Given the spectrum allocation, the operators compete with each other for users present in the system through demand responsive pricing, in the form of an iterative bidding scheme reminiscent of simultaneous ascending auctions. At every iteration, the operators make rate and price offers for each user considering the bandwidth allocated to them and their costs associated with serving the users. The users respond by declaring the probabilities with which they will accept the service offers made. We consider two different objective functions for the SPS to maximize while determining the exact partition of the spectrum resources: (1) the total expected bandwidth utilization, (2) the minimum acceptance probability that a user accepts the offered service. We demonstrate several tradeoffs between these objectives through numerical experiments and illustrate the effect of bandwidth cost on these tradeoffs. Our results also suggest that employing short term allocation of resources could potentially lead to performance gains as opposed to static allocation of resources, especially in regimes where the bandwidth is relatively expensive.}
\end{quotation}

\item \bibentry{Ileri-3}
\begin{quotation}
\footnotesize{\textit{Abstract}: Until recently, the proponents of spectrum commons and the proponents of spectrum property rights had settled, like the armies of the Marne, into their entrenched positions, emerging only to launch periodic and unproductive attacks across a semantic and philosophical no man's land. Recently, however, there have been calls to move beyond these entrenched positions, by developing pragmatic models and solutions that capture some of the benefits of each philosophical position. In this article we cautiously enter that no man's land with two illustrative dynamic spectrum access models. While both of our models retain a bias toward usage of spectrum resources based on a spectrum property rights approach, they also promote dynamic access and short-term dedication of spectrum resources. We call these models dynamic property rights spectrum access (D-Pass) and dynamic commons property rights spectrum access (D-CPass). In this work we focus primarily on engineering issues, proposing the use of a spectrum policy server as a clearinghouse and specifying the spectrum access mechanisms relevant to each model. To demonstrate the useful studies enabled through these models, we present illustrative results via the bandwidth utilization achieved under each model. Our results indicate that both the spectrum access mechanism and market forces will play important roles in determining the resulting bandwidth utilization.}
\end{quotation}

\item \bibentry{Jaya}
\begin{quotation}
\footnotesize{\textit{Abstract}: Hierarchical dynamic spectrum access (DSA) has received the most attention in recent years as the solution for better spectrum utilization. In this paper, on the other hand, we develop a framework for dynamic spectrum leasing (DSL). Power control in hierarchical DSA networks only involves that of controlling secondary user transmissions. Thus, in game theoretic formulations of power control in cognitive DSA networks only secondary users are considered as players of the game. In proposed dynamic spectrum leasing, on the other hand, the primary users are rewarded for allowing secondary users to operate in their licensed spectrum. Thus, in the proposed DSL networks the primary users have an incentive to allow secondary users to access the spectrum whenever possible to the maximum extent. We develop a game theoretic framework for such dynamic spectrum leasing in which primary users actively participate in a non-cooperative game with secondary users by selecting an interference cap on the total interference they willing to tolerate. We establish that the proposed primary-secondary user power control game has a unique Nash equilibrium. Performance of a DSL system based on the proposed game model is compared through simulations under different linear receivers at the secondary base station.}
\end{quotation}

\item \bibentry{Kwon}
\begin{quotation}
\footnotesize{\textit{Abstract}: This paper argues that ''Do auctions raise consumer prices?'' is a misleading question. License fee payment methods, rather than spectrum assignment methods, are key factors that bring forth different market outcomes in the wireless telecommunication industry. This paper analyzes and discusses the effects of three spectrum license fee payment methods-upfront lump-sum fees, royalties, and profit sharing-on economic efficiency, spectrum supply, and government revenue. Royalties create distortions in product and factor markets but can induce the government to increase spectrum supply and encourage firms' investments. A caveat is that the analyses are based on the model assuming monopoly market and information certainty.}
\end{quotation}

\item \bibentry{Lenard}
\begin{quotation}
\footnotesize{\textit{Abstract}: The growth of wireless broadband is a bright spot in the U.S. economy, but a shortage of flexibly licensed spectrum rights could put a crimp on this expansion. Freeing up spectrum from other uses would allow greater expansion of wireless broadband and would bring substantial gains - likely in the hundreds of billions of dollars - for U.S. consumers, businesses, and the federal treasury. In this paper we suggest three methods of making more spectrum available for market-based allocation, from both public and private sources.}
\end{quotation}

\item \bibentry{Li}
\begin{quotation}
\footnotesize{\textit{Abstract}: In this paper, we study the spectrum assignment problem for wireless access networks. Opportunistic spectrum usage is a promising technology. However, it could suffer from the selfish behavior of secondary users. In order to improve opportunistic spectrum usage, we propose to combine the game theory with wireless modeling. Several versions of problems are formalized under different assumptions. We design PTAS or efficient approximation algorithms for each of these problems such that overall social benefit is maximized. Finally, we show how to design a truthful mechanism based on all these algorithms.}
\end{quotation}

\item \bibentry{Licht03}
\begin{quotation}
\footnotesize{\textit{Abstract}: Spectrum is one of the most important and valuable assets of mobile telecommunications operators. According to the options provided in the European Directives, national legislative bodies are in the process of defining the regulatory framework for spectrum trading. While 
the details of the individual regulations in the various member states may differ, spectrum trading will, in any case, confront operators with a new challenge encompassing potential competition, complexity, risks and opportunities. In Germany, a first draft of the new law has been presented for public discussion. In Austria, after a discussion process last year, the new law was supposed to be enacted before the summer of 2003. In the United Kingdom, a draft communication bill is subject to broad and intense discussion. Despite the fact that all concepts are based on the European framework, when analysed in detail, the differences among the various countries are significant. This 
paper describes three national concepts, thereby showing the range of possibilities and considers the competition law criteria that national regulatory authorities (NRA) might utilise when deciding whether to approve individual transactions.}
\end{quotation}

\item \bibentry{Mastro}
\begin{quotation}
\footnotesize{\textit{Abstract}: Licenses for telecommunications services are awarded with a number of side obligations and commitments for the licensee. Under such obligations the licensee is typically not allowed to transfer its license to another operator. Such prohibition may cause heavy inconveniences for customers, so that its removal is strongly advocated and already a reality in many cases. Its removal adds value to the original license and may then constitute a valuable option (the transferability option). A method is here proposed to assess such value, by using the framework of real options. The method is applied in a variety of settings and shows that the value of the option depends superlinearly on the reselling price and the market volatility, and linearly or sub-linearly on the expiry time of the option.}
\end{quotation}
\item \bibentry{Naldi-Aveiro}
\begin{quotation}
\footnotesize{\textit{Abstract}: Dynamic spectrum management makes it possible for the owner of usage rights on some frequency blocks to sublet each of them in real time and for a limited period of time. As a softer implementation with respect to the spot market a two stage assignment is here proposed through the use of options, which give buyers the right to purchase the usage right on a single block and for a timeslot. In the sale of options the primary owner may accomplish an overbooking strategy, which consists in selling more blocks than the available ones and acts as hedging against the risk of unsold blocks. A model for the overbooking strategy is described and evaluated, which takes into account both the value of the option, the correlated decisions taken by the prospective purchasers, and the penalty to be paid to the unsatisfied customers. The dependence of the economical convenience of the overbooking strategy on the relevant parameters (among which the penalty value and the overbooking factor) is shown for a significant range of cases.}
\end{quotation}

\item \bibentry{Naldi-NGI}
\begin{quotation}
\footnotesize{\textit{Abstract}: Mobile Virtual Network Operators (MVNO) operate without owning spectrum usage rights, by leasing spectrum from Mobile Network Operators owning a license. Spectrum leasing may occur through a reservation process that assigns the MVNO the right, but not the obligation, to lease spectrum at a later time. Reservation contracts may be signed with strict guarantees, that provide MVNOs with the certainty of obtaining the needed spectrum, or soft guarantees, which allow the MNO to adopt an overbooking policy and refusing to lease the spectrum, compensating the refused MVNO through the payment of a penalty. In this paper the two kinds of reservation contracts are compared from the viewpoint of the MVNO, by considering the profitability of the two alternatives. The cash flows associated to the two kinds of reservation contracts are determined and employed to compare the alternatives. The probability of overbooking and the expiry time of reservations appear as the major decision factors, while the penalty value has a negligible influence. It is shown that the reservation contract with soft guarantees is to be prefereed for larger values of the expiry time and for the lower values of the overbooking probability.}
\end{quotation}

\item \bibentry{Naldi-CTTE}
\begin{quotation}
\footnotesize{\textit{Abstract}: With spectrum trading the owners of usage rights on wireless spectrum may transfer them temporarily to other users. A secondary market for spectrum may be implemented through the use of reservation mechanisms, by which prospective consumers of spectrum portions may first book them and decide later whether to actually buy the usage rights. A possible strategy for the owners of usage rights is to practise overbooking, i.e., to allow for more reservations than what they could sell (and then pay a penalty if they cannot satisfy the requests). A relevant problem for the owners of usage rights is to set the price for reservations. An algorithm is proposed here to price reservations in the presence of overbooking. The algorithm is based on the iterative use of the Cox-Ross-Rubinstein approach to price financial \textit{call} options. The algorithm is then applied to a number of cases and the dependence of the price on the overbooking probability and the penalty is analysed.}
\end{quotation}

\item \bibentry{Mayo}
\begin{quotation}
\footnotesize{\textit{Abstract}: Despite the potentially critical nature of secondary markets in maintaining efficient spectrum and wireless markets, research has to date has focused primarily on the Federal Communication Commission’s rules for initial distribution of spectrum. To redress this lacuna, we first examine the evolution of conceptual and policy developments directed toward secondary spectrummarkets. Then, we seek to move beyond those efforts to empirically document the development of secondary spectrum activity in the United States and the relationship of that development to the evolving policy toward such markets. We categorize and explore different types of secondary spectrum markets. Then, by drawing on a database of every spectrum license transaction since 1994, we explore the depth and breadth of spectrum trading in secondary markets. We find that the FCC has radically reduced the time it takes to approve trades, making the system more akin to notification than to approval. We also find that a large amount of spectrum changes hands each year. While these conclusions do not necessarily imply that secondary markets work efficiently, they do show that policy efforts to facilitate and energize the growth of secondary spectrummarkets are bearing considerable fruit.}
\end{quotation}

\item \bibentry{Mohsenian}
\begin{quotation}
\footnotesize{\textit{Abstract}: In a multi-hop wireless access network, where each node is an independent self-interested commercial entity, pricing is helpful not only to encourage collaboration but also to utilize the network resources efficiently. In this paper, we propose a market-based model with two-fold pricing (TFP) for wireless access networks. In our model, the relay-pricing is used to encourage nodes to forward packets for other nodes. Each node receives a payment for the relay service that it provides. We also consider interference-pricing to leverage optimal resource allocation. Together, the relay and interference prices incorporate both cooperative and competitive interactions among the nodes. We prove that TFP guarantees positive profit for each individual wireless node for a wide range of pricing functions. The profit increases as the node forwards more packets. Thus, the cooperative nodes are well rewarded. We then determine the relay and interference pricing functions such that the network social welfare and the aggregate network utility are maximized. Simulation results show that, compared to two recently proposed single-fold pricing models, where only the relay or only the interference prices are considered, our proposed TFP scheme significantly increases the total network profit as well as the aggregate network throughput. TFP also leads to more fair revenue sharing among the wireless relay nodes.}
\end{quotation}

\item \bibentry{Mutlu}
\begin{quotation}
\footnotesize{\textit{Abstract}: Recent deregulation initiatives enable cellular providers to sell excess spectrum for secondary usage. In this paper, we investigate the problem of optimal spot pricing of spectrum by a provider in the presence of both nonelastic primary users, with long-term commitments, and opportunistic, elastic secondary users. We first show that optimal pricing can be formulated as an infinite horizon average reward problem and solved using stochastic dynamic programming. Next, we investigate the design of efficient single pricing policies. We provide numerical and analytical evidences that static pricing policies do not perform well in such settings (in sharp contrast to settings where all the users are elastic). On the other hand, we prove that deterministic threshold pricing achieves optimal profit amongst all single-price policies and performs close to global optimal pricing. We characterize the profit regions of different pricing policies, as a function of the arrival rate of primary users. Under certain reasonable assumptions on the demand function, we prove that the profit region of threshold pricing is optimal and independent of the specific form of the demand function, and that it includes the profit region of static pricing. In addition, we show that the profit function of threshold pricing is unimodal in price. We determine a restricted interval in which the optimal threshold lies. These properties enable very efficient computation of the optimal threshold policy, which is far faster than that of the global optimal policy.}
\end{quotation}

\item \bibentry{Neely}
\begin{quotation}
\footnotesize{\textit{Abstract}: We consider an ad-hoc wireless network operating within a free market economic model. Users send data over a choice of paths, and scheduling and routing decisions are updated dynamically based on time varying channel conditions, user mobility, and current network prices charged by intermediate nodes. Each node sets its own price for relaying services, with the goal of earning revenue that exceeds its time average reception and transmission expenses. We first develop a greedy pricing strategy that maximizes social welfare while ensuring all participants make non-negative profit. We then construct a (non-greedy) policy that balances profits more evenly by optimizing a profit fairness metric. Both algorithms operate in a distributed manner and do not require knowledge of traffic rates or channel statistics. This work demonstrates that individuals can benefit from carrying wireless devices even if they are not interested in their own personal communication.}
\end{quotation}

\item \bibentry{Niyato}
\begin{quotation}
\footnotesize{\textit{Abstract}: We consider the problem of spectrum trading with multiple licensed users (i.e., primary users) selling spectrum opportunities to multiple unlicensed users (i.e., secondary users). The secondary users can adapt the spectrum buying behavior (i.e., evolve) by observing the variations in price and quality of spectrum offered by the different primary users or primary service providers. The primary users or primary service providers can adjust their behavior in selling the spectrum opportunities to secondary users to achieve the highest utility. In this paper, we model the evolution and the dynamic behavior of secondary users using the theory of evolutionary game. An algorithm for the implementation of the evolution process of a secondary user is also presented. To model the competition among the primary users, a noncooperative game is formulated where the Nash equilibrium is considered as the solution (in terms of size of offered spectrum to the secondary users and spectrum price). For a primary user, an iterative algorithm for strategy adaptation to achieve the solution is presented. The proposed game-theoretic framework for modeling the interactions among multiple primary users (or service providers) and multiple secondary users is used to investigate network dynamics under different system parameter settings and under system perturbation.}
\end{quotation}

\item \bibentry{Niyato-2}
\begin{quotation}
\footnotesize{\textit{Abstract}: We consider the problem of hierarchical bandwidth sharing in dynamic spectrum access (or cognitive radio) environment. In the system model under consideration, licensed service (i.e., primary service) can share/sell its available bandwidth to an unlicensed service (i.e., secondary service), and again, this unlicensed service can share/sell its allocated bandwidth to other services (i.e., tertiary and quaternary services). We formulate the problem of hierarchical bandwidth sharing as an interrelated market model used in microeconomics for which a multiple-level market is established among the primary, secondary, tertiary, and quaternary services. We use the concept of demand and supply functions to obtain the equilibrium at which all the services are satisfied with the amount of allocated bandwidth and the price. These demand and supply functions are derived based on the utility of the connections using the different services (i.e., primary, secondary, tertiary, and quaternary services). For distributed implementation of the hierarchical bandwidth sharing model in a system in which global information is not available, iterative algorithms are proposed through which each service adapts its strategies to reach the equilibrium. The system stability condition is analyzed for these algorithms. Finally, we demonstrate the application of the proposed model to achieve dynamic bandwidth sharing in an integrated WiFi-WiMAX network.}
\end{quotation}

\item \bibentry{Niyato-3}
\begin{quotation}
\footnotesize{\textit{Abstract}: "Cognitive radio" is an emerging technique to improve the utilization of radio frequency spectrum in wireless networks. In this paper, we consider the problem of spectrum sharing among a primary user and multiple secondary users. We formulate this problem as an oligopoly market competition and use a noncooperative game to obtain the spectrum allocation for secondary users. Nash equilibrium is considered as the solution of this game. We first present the formulation of a static game for the case where all secondary users have the current information of the adopted strategies and the payoff of each other. However, this assumption may not be realistic in some cognitive radio systems. Therefore, we consider the case of bounded rationality in which the secondary users gradually and iteratively adjust their strategies based on the observations on their previous strategies. The speed of adjustment of the strategies is controlled by the learning rate. The stability condition of the dynamic behavior for this spectrum sharing scheme is investigated. The numerical results reveal the dynamics of distributed dynamic adaptation of spectrum sharing strategies.}
\end{quotation}

\item \bibentry{Niyato-4}
\begin{quotation}
\footnotesize{\textit{Abstract}: The emerging IEEE 802.22-based wireless regional area network technology will use the same radio spectrum currently allocated for TV service. This standard will use the concept of cognitive radio based on dynamic spectrum access to provide wireless access services in a large coverage area. A brief overview of the current state of the IEEE 802.22 standard is provided with a particular emphasis on the spectrum management (i.e., spectrum sensing and dynamic spectrum access) in this standard. Key research issues related to spectrum trading among TV broadcasters, WRAN service providers, and IEEE 802.22-based cognitive radio users are identified. To this end, a hierarchical spectrum trading model is presented to analyze the interaction among WRAN service providers, TV broadcasters, and WRAN users. In this model a double auction is established among multiple TV broadcasters and WRAN service providers who sell and buy the radio spectrum (i.e., TV bands), respectively. Again, multiple WRAN service providers compete with each other by adjusting the service price charged to WRAN users. We propose a joint spectrum bidding and service pricing model for WRAN service providers to maximize their profits. A non-cooperative game is formulated to obtain the solution in terms of the number of TV bands and the service price of a service provider. Numerical results are presented on the performance of this joint spectrum bidding and pricing model.}
\end{quotation}

\item \bibentry{Niyato-5}
\begin{quotation}
\footnotesize{\textit{Abstract}: Cognitive radio has emerged as a new design paradigm for the next-generation wireless networks. Cognitive radio networks are designed based on the concept of dynamic spectrum sharing where cognitive radio users can opportunistically share the radio spectrum. For example, in case of vertical spectrum sharing, a spectrum owner (or primary user/service) can share (or sell) its licensed spectrum with (to) other users (i.e., secondary users/services). The economics of spectrum sharing, which is referred to as spectrum trading, is the focus of this article. The objective of spectrum trading is to maximize the revenue of the spectrum owner, and at the same time enhance the satisfaction of the cognitive radio users. We first discuss the different network architectures and protocol behaviors for dynamic spectrum sharing as well as the spectrum sharing models. Then the scope of spectrum trading is discussed in the context of different spectrum sharing models. The primary research issues related to spectrum trading in cognitive radio networks are outlined, and the possible solution approaches are discussed. To this end, we introduce a market-equilibrium-based spectrum trading mechanism that uses spectrum demand and supply of the primary and secondary users, respectively. Since spectrum supply is stochastic in nature, a distributed and adaptive learning algorithm is used for the secondary users to estimate spectrum price and adjust the spectrum demand accordingly so that the market equilibrium can be reached.}
\end{quotation}

\item \bibentry{Olafsson}
\begin{quotation}
\footnotesize{\textit{Abstract}: As the heterogeneity of wireless access technologies increases, dynamic allocation and utilisation of spectrum become ever more important. The traditional rigid allocation of spectrum for technology-specific usage is not suitable for the increasingly dynamic demand driven by the continuous emergence of technologies providing new services with different quality of service requirements. New spectrum management techniques and increasingly flexible spectrum usage rights are therefore called for. We discuss the limitations of present spectrum management techniques and explore some new alternatives including spectrum trading and opportunistic spectrum access.}
\end{quotation}

\item \bibentry{Peha-1}
\begin{quotation}
\footnotesize{\textit{Abstract}: As market-based reform sweeps telecommunications industries around the world, it is a good time to reevaluate the spectrum management policies which govern wireless industries ranging from broadcast television to satellite communications. Most countries have been using a central planning approach to spectrum management, but there are many alternatives with varying degrees of flexibility and market-based incentives. This paper provides a survey of spectrum management approaches, addressing methods of determining how spectrum can be used, which commercial entities can use it, and how governments can manage their own spectrum. It identifies some of the crucial choices to be made, and summarizes advantages and disadvantages of each.}
\end{quotation}

\item \bibentry{Peha-2}
\begin{quotation}
\footnotesize{\textit{Abstract}: Many complain about severe spectrum shortage. The shortage comes from outdated spectrum policies that allows for little sharing. Regulators have granted licenses that offer exclusive access to the spectrum. When these licensees are not transmitting, the spectrum sits idle. A new technology regarding spectrum shortage enables more spectrum sharing that unleashes innovative products and services, provided that we adopt the appropriate spectrum policies. Two camps are pushing for extreme reform, one for "property rights" and the other for "spectrum commons". This article presents concepts underlying the "property" and "commons" debate, clarifies options for spectrum reform, and describes the trade-offs of spectrum sharing.}
\end{quotation}

\item \bibentry{Peha-3}
\begin{quotation}
\footnotesize{\textit{Abstract}: Traditionally, interference protection is guaranteed through a policy of spectrum licensing, whereby wireless systems get exclusive access to spectrum. This is an effective way to prevent interference, but it leads to highly inefficient use of spectrum. Cognitive radio along with software radio, spectrum sensors, mesh networks, and other emerging technologies can facilitate new forms of spectrum sharing that greatly improve spectral efficiency and alleviate scarcity, if policies are in place that support these forms of sharing. On the other hand, new technology that is inconsistent with spectrum policy will have little impact. This paper discusses policies that can enable or facilitate use of many spectrum-sharing arrangements, where the arrangements are categorized as being based on coexistence or cooperation and as sharing among equals or primary-secondary sharing. A shared spectrum band may be managed directly by the regulator, or this responsibility may be delegated in large part to a license-holder. The type of sharing arrangement and the entity that manages it have a great impact on which technical approaches are viable and effective. The most efficient and cost-effective form of spectrum sharing will depend on the type of systems involved, where systems under current consideration are as diverse as television broadcasters, cellular carriers, public safety systems, point-to-point links, and personal and local-area networks. In addition, while cognitive radio offers policy-makers the opportunity to improve spectral efficiency, cognitive radio also provides new challenges for policy enforcement. A responsible regulator will not allow a device into the marketplace that might harm other systems. Thus, designers must seek innovative ways to assure regulators that new devices will comply with policy requirements and will not cause harmful interference.}
\end{quotation}

\item \bibentry{Pogorel03}

\item \bibentry{Sengupta}
\begin{quotation}
\footnotesize{\textit{Abstract}: The concept of dynamic spectrum access will allow the radio spectrum to be traded in a market like scenario allowing wireless service providers (WSPs) to lease chunks of spectrum on a short-term basis. Such market mechanisms will lead to competition among WSPs where they not only compete to acquire spectrum but also attract and retain users. Currently, there is little understanding on how such a dynamic trading system will operate so as to make the system feasible under economic terms. In this paper, we propose an economic framework that can be used to guide i) the dynamic spectrum allocation process and ii) the service pricing mechanisms that the providers can use. We propose a knapsack based auction model that dynamically allocates spectrum to the WSPs such that revenue and spectrum usage are maximized. We borrow techniques from game theory to capture the conflict of interest between WSPs and end users. A dynamic pricing strategy for the providers is also proposed. We show that even in a greedy and non-cooperative behavioral game model, it is in the best interest of the WSPs to adhere to a price and channel threshold which is a direct consequence of price equilibrium. Through simulation results, we show that the proposed auction model entices WSPs to participate in the auction, makes optimal use of the spectrum, and avoids collusion among WSPs. We demonstrate how pricing can be used as an effective tool for providing incentives to the WSPs to upgrade their network resources and offer better services.}
\end{quotation}

\item \bibentry{Stine}
\begin{quotation}
\footnotesize{\textit{Abstract}: Spectrum management is the process of deciding how radio frequency (RF) spectrum may be used in a geographical region and who may use it. Traditionally, spectrum management has been executed as an administrative and political process with the intent of making lasting decisions. Its lack of responsiveness and resolution causes much spectrum to lay fallow since most users rarely need spectrum continuously and ubiquitously. In this paper, we propose an alternative spectrum management approach that enables management at a greater temporal and spatial resolution using networks and wireless ad hoc and mesh networking technologies. Three different spectrum management ideas are described. The Synchronous Collision Resolution (SCR) MAC protocol enables a strict arbitration of spectrum access based on spectrum rights thus enabling a hierarchy of networks in the same spectrum that always guarantees the primary rights holder precedence. Second, it autonomously manages the use of an arbitrary number of channels in the same network. The third and most exciting idea is a new fast command and control model for spectrum management. An underlying ad hoc network built using the Nodes State Routing* (NSR) protocol is used to track and manage the use of spectrum of attached RF emitters. NSR tracks the state of the network by collecting and disseminating the states of the nodes. These states can include relevant information on the spectrum these nodes are using and are observing others use. Thus the network supports tracking and monitoring spectrum use spatially in near real time. Spectrum management utilities built on top of the network could allow users and spectrum managers to rapidly negotiate the use of spectrum and assist spectrum managers in identifying unused spectrum and emitters causing harmful interference. We conclude with proposed standardization and regulatory changes to make this feasible.}
\end{quotation}

\item \bibentry{Stumpf}
\begin{quotation}
\end{quotation}

\item \bibentry{Toka}
\begin{quotation}
\footnotesize{\textit{Abstract}: The emergence of novel radio techniques enables the application of advantageous revolutionary spectrum policies. An important body of research has appeared about possible frequency management schemes, but none of them proposes solutions that meet every related criteria. In this paper we present our work on dynamic spectrum allocation and pricing that offers a distributed mechanism design, well-suited to practical employment issues. Our model handles interference effects without any restricting assumptions, provides universal scalable and incentive-compatible allocation and pricing mechanisms. We provide both analytical and numerical evaluation of the proposed framework, and in either case we prove this latter to be a suitable approach to efficient and flexible spectrum utilization.}
\end{quotation}

\item \bibentry{Valletti01}
\begin{quotation}
\footnotesize{\textit{Abstract}: This paper argues that the current centralised model of spectrum management is highly inefficient and should be replaced with decentralised solutions. The current model suppresses competitive entry, blocks efficient spectrum use, and insulates old technologies from innovative challenge. In the new system, the default rule should endow operators with the highest flexibility, leaving the regulator to monitor the proper working of competition rather than deciding who does what.}
\end{quotation}

\item \bibentry{Wang}
\begin{quotation}
\footnotesize{\textit{Abstract}: Cognitive radio is emerging as a promising technique to improve the utilization of the radio frequency spectrum. In this paper, we consider the problem of spectrum sharing among primary (or "licensed") users (PUs) and secondary (or "unlicensed") users (SUs). We formulate the problem based on bandwidth auction, in which each SU makes a bid for the amount of spectrum and each PU may assign the spectrum among the SUs by itself according to the information from the SUs without degrading its own performance. We show that the auction is a noncooperative game and that Nash equilibrium (NE) can be its solution. We first consider a single-PU network to investigate the existence and uniqueness of the NE and further discuss the fairness among the SUs under given conditions. Then, we present a dynamic updating algorithm in which each SU achieves NE in a distributed manner. The stability condition of the dynamic behavior for this spectrum-sharing scheme is studied. The discussion is generalized to the case in which there are multiple PUs in the network, where the properties of the NE are shown under appropriate conditions. Simulations were used to evaluate the system performance and verify the effectiveness of the proposed algorithm.}
\end{quotation}

\item \bibentry{Willkomm}
\begin{quotation}
\footnotesize{\textit{Abstract}: Dynamic spectrum access approaches, which propose to opportunistically use underutilized portions of licensed wireless spectrum such as cellular bands, are increasingly being seen as a way to alleviate spectrum scarcity. However, before DSA approaches can be enabled, it is important that we understand the dynamics of spectrum usage in licensed bands. Our focus in this article is the cellular band. Using a unique dataset collected inside a cellular network operator, we analyze the usage in cellular bands and discuss the implications of our results on enabling DSA in these bands. One of the key aspects of our dataset is its scale-it consists of data collected over three weeks at hundreds of base stations. We dissect this data along different dimensions to characterize if and when spectrum is available, develop models of primary usage, and understand the implications of these results on DSA techniques such as sensing.}
\end{quotation}

\item \bibentry{WuWang}
\begin{quotation}
\footnotesize{\textit{Abstract}: Dynamic spectrum access (DSA), enabled by cognitive radio technologies, has become a promising approach to improve efficiency in spectrum utilization, and the spectrum auction is one important DSA approach, in which secondary users lease some unused bands from primary users. However, spectrum auctions are different from existing auctions studied by economists, because spectrum resources are interference-limited rather than quantity-limited, and it is possible to award one band to multiple secondary users with negligible mutual interference. To accommodate this special feature in wireless communications, in this paper, we present a novel multi-winner spectrum auction game not existing in auction literature. As secondary users may be selfish in nature and tend to be dishonest in pursuit of higher profits, we develop effective mechanisms to suppress their dishonest/collusive behaviors when secondary users distort their valuations about spectrum resources and interference relationships. Moreover, in order to make the proposed game scalable when the size of problem grows, the semi-definite programming (SDP) relaxation is applied to reduce the complexity significantly. Finally, simulation results are presented to evaluate the proposed auction mechanisms, and demonstrate the complexity reduction as well.}
\end{quotation}

\item \bibentry{Xavier}
\begin{quotation}
\footnotesize{\textit{Abstract}: \textbf{Purpose} - An aspect of spectrum reform receiving increasing attention is the introduction of secondary markets for spectrum in order to enable more flexibility to reassign unused and underused spectrum to users that will use it more efficiently. This paper proposes to focus on the policy issues relating to the development of well-functioning secondary markets for spectrum.\\
\textbf{Design/methodology/approach} - The paper reviews developments in the debate over secondary markets for spectrum. It draws together key elements from the academic literature, various government and government-commissioned reports, and the practical experience of the few countries that have already introduced spectrum trading. There is considerable focus on concerns and potential costs relating to the introduction of spectrum trading and liberalisation. This has a constructive aim - to draw attention to the need to address such concerns in order to facilitate the development of spectrum trading.\\
\textbf{Findings} - While there is a persuasive case for spectrum trading, countries have been slow to introduce it because of a number of concerns. This paper identifies these concerns and the regulatory framework/policies needed to address them.\\
\textbf{Originality/value} - The paper distils the policy issues in the debate over spectrum trading and identifies the role that regulators will need to play in the introduction, facilitation and regulation of secondary markets for spectrum.}
\end{quotation}

\item \bibentry{Xing}
\begin{quotation}
\footnotesize{\textit{Abstract}: We explore the price dynamics in a competitive market consisting of spectrum agile network service providers and users. Here, multiple self interested spectrum providers operating with different technologies and costs compete for potential customers. Different buyers or consumers may evaluate the same seller differently depending on their applications, operating technologies and locations. Two different buyer populations, the quality-sensitive and the price-sensitive are investigated, and the resulting collective price dynamics are studied using a combination of analysis and simulations. Various scenarios are considered regarding the nature and accuracy of information available to the sellers. A myopically optimal strategy is studied when full information is available, while a stochastic learning based strategy is considered when the information is limited. Cooperating groups may be formed among the sellers which will in-turn influence the group profit for those participants. Free riding phenomenon is observed under certain circumstances.}
\end{quotation}

\item \bibentry{Yu}
\begin{quotation}
\footnotesize{\textit{Abstract}: Spectrum under-utilization is one of the bottlenecks of the development of wireless communication, and dynamic spectrum access (DSA) is envisioned as a novel mechanism to solve the problem of spectrum scarcity. Spectrum auction has been recognized as an effective way to achieve DSA, wherein the primary spectrum owner (PO) acts as an auctioneer who has free channels and is willing to sell them for additional revenue, and the secondary user (SU) acts as a bidder who is willing to buy a channel from POs for its service. In this paper, we adopt a progressive spectrum auction named MAP, which has been proved optimal and incentive compatible in DSA networks with distributed POs and SUs. However, in MAP, the profit of POs is not maximized under the equilibrium point due to the scarcity of SUs' private information known by POs. We propose an information sharing mechanism, in which the POs exchange their local information with each other. We show analytically that, allowing information sharing, each PO is able to learn the private information of SUs and increase its profit accordingly. Long term profit acts as the incentive for information sharing that all the POs automatically reveal the true information when they are aware of this. It is notable that information sharing doesn't affect social optimality. Simulation shows the increase of POs' profits in the sense of long term interests.}
\end{quotation}

\end{itemize}

\end{document}